\documentclass{aa}
\usepackage{graphicx}
\usepackage{natbib}
\usepackage{txfonts}

\usepackage{fix2col}

\bibpunct{(}{)}{;}{a}{}{,}

\newcommand\changed{}

\begin{document}

\title{Pulse-to-pulse intensity modulation and drifting subpulses in
   recycled pulsars}
\author{R.~T. Edwards\inst{1} \and
   B.~W. Stappers\inst{2,1}} 
\titlerunning{Pulse intensity modulation in recycled pulsars} 
\institute{Astronomical Institute ``Anton
   Pannekoek'', University of Amsterdam, Kruislaan 403, 1098 SJ
   Amsterdam, The Netherlands \and Stichting ASTRON, Postbus 2, 7990
   AA Dwingeloo, The Netherlands } \offprints{R.~T. Edwards,
   \email{redwards@astro.uva.nl}}

\abstract{ We report the detection of pulse-to-pulse periodic
intensity modulations, in observations of recycled pulsars. Even
though the detection of individual pulses was generally not possible
due to their low flux density and short duration, through the
accumulation of statistics over sequences of $10^5$--$10^6$ pulses we
were able to determine the presence and properties of the
pulse-to-pulse intensity variations of six pulsars.  In most cases we
found that the modulation included a weak, broadly quasi-periodic
component. For two pulsars the sensitivity was high enough to
ascertain that the modulation phase apparently varies systematically
across the profile, indicating that the modulation appears as drifting
subpulses. We detected brighter than average individual pulses in
several pulsars, with energies up to 2--7 times higher than the mean,
similar to results from normal pulsars. We were sensitive to giant
pulses of a rate of occurrence equal to (and in many instances much
lower than) that of PSR B1937+21 at 1400~MHz ($\sim 30$ times lower
than at 430~MHz), but none were detected, indicating that the
phenomenon is rare in recycled pulsars. 
}

\maketitle

\section{Introduction}
Pulsar science is perhaps unusual in astronomy, in that progress in
the theoretical understanding of the emission mechanism is not limited
by a lack of data, but rather by an overabundance of complex
observable pheonomenology. Pulsar signals are seen to vary on every
observable timescale from nanoseconds (e.g. \citealt{hkwe03}) to
decades (e.g. \citealt{wt02}), with variations associated with such
phenomena as microstructure, subpulses, periodic/drifting subpulses,
giant pulses, pulse nulling, mode changing, interstellar scintillation
and geodetic spin precession.  All of these phenomena have been
important in shaping ideas concerning the radio pulsar emission
process, however to date a complete explanation of radio pulsar
emission remains elusive.

This discovery of PSR B1937+21 \citep{bkh+82} posed further challenges
due to its very short spin period of $1.56$~ms. As more ``millisecond
pulsars'' (MSPs) were discovered, the rapid spin and low inferred
surface magnetic field strength grew to be understood as the end
product of accretion from a binary companion, ``recycling'' pulsars that
had earlier evolved into radio silence (e.g. \citealt{pk94}).
However, the question of how the radio emission mechanism is related to
that of ordinary pulsars remained open. With a difference of 2--3
orders of magnitude in rotation period and 3--4 orders of magnitude in
magnetic field strength, a common emission mechanism might not be
expected, yet a good argument can be made for this through the
comparison of (polarimetric) average pulse profile morphologies and
their frequency dependence (\citealt{kx00} and references
therein). 

As noted above, the true pulsar signal varies on all timescales and
much has been learned about ``ordinary'' pulsars from studying aspects
of the signal other than the basic average profile. This has not been
the case with MSPs because their typically lower flux densities and
shorter pulses make the detection of individual pulses
difficult. An exception has been the study of so-called ``giant''
pulses, in PSR B1937+21 \citep{wcs84,bac95,cstt96,kt00,viv02,ps03},
and PSR B1821$-$24 \citep{rj01}. Studies of the properties of pulses
of normal intensity have thus far been limited to three pulsars.
\citet{sb95} used the 305-m Arecibo dish to obtain sensitivity to
individual pulses of PSR B1534+12 of average and above-average energy
and found that their distribution in energy was similar to that of
some ordinary pulsars. On the basis of the detection of pulses of
above-average intensity, both \citet{jak+98} and \citet{vam98} found
that the properties of the brightest known MSP, PSR J0437$-$4715, were
similar to those of ordinary pulsars, although to good significance
neither nulling nor the presence of preferred timescales within
individual pulses (e.g. microstructure, subpulses) was detected. Both
studies report the detection of quasi-periodic pulse-to-pulse
intensity modulations with a period of $\sim 4$ pulses, and
\citet{vam98} report (we believe potentially erroneously; see
Sect. \ref{sec:discussion}) that the modulation is not associated with
drifting subpulses, as it often is in ordinary pulsars. Finally,
\citet{jap01} used data with an average single-pulse signal-to-noise
(S/N) ratio less than one to show that on a statistical basis and with
the exception of giant pulses, the emission of PSR B1937+21 is
extremely stable, showing to high significance a complete lack of
pulse-to-pulse variations and no difference between the shape of
individual pulses and that of the average profile.

As shown by \citet{jap01}, poor single-pulse S/N does not eliminate
the possibility of obtaining useful information regarding single
pulses. The average profile can of course be measured in such cases,
but there is no reason why collection of statistics cannot be extended
beyond the first moment to include higher-order moments and
correlations. Of primary interest are the second
moment\footnote{Since they worked from voltage samples instead of
intensity samples, \citet{jap01} referred to this as the fourth
moment.} (giving access to the modulation index), and the second order
correlations. A variety of useful correlation statistics can be
derived; \citet{jap01} integrated the single-pulse autocorrelation
function (ACF) to show that every pulse appeared to be identical to
the average profile.  We extend this approach here to include the
Longitude-Resolved Fluctuation Spectrum (LRFS; \citealt{bac70b}), the
Two-Dimensional Fluctuation Spectrum (2DFS; \citealt{es02}) and the
Longitude-Resolved Cross-Correlation Function (LRCCF;
\citealt{pop86}).
Applying these techniques to archival observations made at the
Westerbork Synthesis Radio Telescope (WSRT),
we have detected and characterised pulse modulation behaviour
in several recycled pulsars.

\section{Observations and Methods of Analysis}
\subsection{Observations}
The data used in this project were selected from archival data taken
at WSRT. For many pulsars numerous observations were available,
allowing us to take advantage of fortuitous scintillation conditions
for substantial enhancements in sensitivity.
For all observations, dual linear polarization signals from fourteen
25-m dishes were added using previously determined phase and gain
factors and the resultant signal was digitally processed by the PuMa
pulsar backend to form a two-dimensional array of total power samples,
as a function of time and radio frequency; for details see
\citet{vkv02}. In offline processing, frequency channels containing
periodic interference were flagged and a de-dispersed time series was
produced by combining remaining channels, aligned using previously
published dispersion measures. In Table \ref{tab:obs} we list
parameters of the observations used, including the date of observation
as a Modified Julian Day (MJD), the observation duration ($T_{\rm
obs}$) in seconds, the centre frequency ($\nu$) and total bandwidth
($\Delta\nu$) in megahertz, the number of frequency channels used, the
resultant dispersion smearing ($\tau_{\rm smear}$, near the center
frequency) and output sample interval ($\tau_{\rm samp}$) in
microseconds, and the number of pulses recorded . The rightmost four
columns refer to the sensitivity, see Sect. \ref{sec:nondet}.

\begin{table*}
\begin{center}
\caption{Observation parameters and sensitivity}
\begin{tabular}{rrrrrrr@{.}lr@{.}lrr@{.}lr@{.}lr@{.}lr}
Name& Date  & $T_{\rm obs}$ & $\nu$ & $\Delta\nu$  
  & $N_{\rm chan}$ & \multicolumn{2}{c}{$\tau_{\rm smear}$} 
  & \multicolumn{2}{c}{$\tau_{\rm samp}$}
  & $N_{\rm pulses}$
  & \multicolumn{2}{c}{$\sigma_{\rm g}$} 
  & \multicolumn{2}{c}{$\sigma_{\rm m}$}
  & \multicolumn{2}{c}{$\sigma_{\rm c}$} 
  & $E_{\rm min}$ \\
(PSR) & (MJD) &  (s)          & (MHz) &    (MHz)     &   
  & \multicolumn{2}{c}{($\mu$s)} & \multicolumn{2}{c}{($\mu$s)} 
  & ($10^3$)
  & \multicolumn{6}{c}{} 
  & ($<E>$) \\
\\
\hline
J0034$-$0534 & 52216.1 & 1800 & 328  & 10 & 512 & 63&2 & 102&4 
             & 775 &2&0 & 1&7 & 0&11 & 20 \\
J0218+4232   & 52462.4 & 1800 & 328  & 10 & 1024 & 140&7 & 102&4
             & 959 & 3&6 & 3&0 & 0&19 & 26\\
J0613$-$0200 & 52190.1 & 9000 & 1380 & 80 & 512 & 19&1& 51&2
             & 2914 & 3&1 & 2&6 & 0&16 & 19 \\
J1012+5307   & 52145.7 & 1800 & 840  & 80 & 512 & 19&7 & 102&4 
             & 342 & 0&37 & 0&31 & 0&019 & 3.3 \\
             & 52392.8 & 1800 & 1380 & 80 & 512 & 4&5  & 51&2 
             & 342 & 0&2 & 0&17 & 0&011 & 0.78\\
J1022+1001   & 52004.8 & 7200 & 328 & 10 & 512 & 47&0 & 102&4
             & 433 & 0&29 & 0&24 & 0&015 & 1.5 \\
             & 52004.6 & 7200 & 840 & 80 & 64 & 22&4 & 51&2
             & 433 & 0&55 & 0&47 & 0&029 & 2.2 \\
             & 52086.6 & 3600 & 1380 & 80 & 512 & 5&1 & 102&4
             & 433 & 0&21 & 0&17 & 0&011 & 0.62 \\
J1518+4904   & 52101.7 & 3600 & 1390 & 60 & 384 & 5&6 & 204&8
             & 85 & 0&096& 0&081 & 0&0057$^2$ & 0.65\\
J1643$-$1224 & 52242.5 & 1800 & 1380 & 80 & 512 & 30&8 & 102&4
             & 368 & 1&0 & 0&85 & 0&053 & 12\\
J1713+0747   & 52544.7 & 7200 &  840 & 80 & 512 & 35&0 & 51&2 
             & 1558 & 0&54 & 0&46 & 0&028 & 8.7 \\
             & 52515.7 & 1800 & 1190 & 80 & 512 & 12&3 & 51&2
             & 376 & 0&12 & 0&10 & 0&0064 & 1.8\\
             & 52537.7 & 3600 & 1700 & 40 & 256 & 4&2  & 51&2
             & 770 & 0&11 & 0&092 & 0&0058 & 2.1\\
             & 52572.5 & 3600 & 2240 & 80 & 512 & 1&8  & 51&2 
             & 770 & 0&27 & 0&23 & 0&014 & 5.2\\
J1918$-$0642 & 52560.8 & 1800 & 1380 & 80 & 512 & 13&1 & 102&4
             & 235 & 0&58 & 0&49 & 0&030 & 5.8 \\
B1937+21     & 52403.3 & 900 & 1380 & 80 & 512 & 35&0 & 25&6 
             & 576 & 0&31 & 0&26 & 0&016 & 5.3 \\
J2145$-$0750 & 52223.7 & 1500 &  860 & 40 & 256 & 19&7 & 102&4 
             & 91 & 0&38 & 0&32 & 0&020 & 1.2 \\
             & 52232.9 & 3600 & 1380 & 80 & 512 & 4&4  & 102&4
             & 222 & 0&21 & 0&18 & 0&011 & 0.92 \\
\hline
\end{tabular} \\
$^1$: $M_{\rm max} = 24576$
$^2$: $M_{\rm max} = 81920$
\label{tab:obs}
\end{center}
\end{table*}

\subsection{Production of a Longitude/Time-dependent Array}
For many techniques of single pulse analysis including those used in
this work, the data need to be treated as a sequence of sampled
pulses, with a consistent sampling lattice within each pulse.  The
simplest approach to producing a such two-dimensional (i.e. longitude-
and time-resolved) representation of the pulse sequence is to bin each
sample according to according its pulse phase, as calculted using an
ephemeris (for which purpose we used the {\sc TEMPO} software 
package\footnote{http://pulsar.princeton.edu/tempo/}).
 When the time resolution is close to the interval
corresponding to the desired longitude resolution, a problem arises
with this method. In the simplest case, where the longitude bin width
is chosen to correspond to one sample interval (at some point during
the observation), each pulse as it appears in the binned sequence is
effectively shifted by an amount corresponding to the offset between
each sample and the bin centers. This offset advances (modulo one bin)
by a constant amount each pulse due to the fact that the apparent
pulse period does not (in general) equal an integer number of sample
intervals. The pulse shape distortion (longitude shift) is thus
periodic from pulse to pulse, and gives rise to a sequence of
harmonics in the LRFS. This problem was encountered by
\citet{vam98} in their analysis of data from PSR J0437$-$1715, but
no attempt was made to remove it.

In this work we avoided the effect by compensating for the shift in
each pulse using a frequency-domain time shift.  Formally, we may view
the time series as a sampled signal that represents the source
intensity signal, convolved with a function representing the applied
integration. By using a sequence of samples that is offset from the
required set of bin centers (longitude samples) by some fraction of a
sample $\epsilon \in [-0.5,0.5]$, and assigning those samples to the
nearest bin, we arrive at a version of the pulse signal that is
apparently longitude-shifted by $\epsilon$ samples. This can be simply
corrected by convolution with an offset sinc function, which by virtue
of Fast Fourier Transform algorithms is efficiently performed by
multiplying the Discrete Fourier Transform (DFT) of the pulse by
$\Delta(\nu_{\rm l})=\exp(2\pi i \epsilon\nu_{\rm l})$ and taking
the inverse DFT\footnote{Since this performs a cyclical convolution,
it is necessary to discard some samples from the ends of the
sequence. This is most noticable in the presence of strong
interference, when the apparent step between the last sample and the
first can cause ringing at the Nyquist freqency when the sampling
lattice is offset by this process. We found that discarding 32 samples
from each end of the result was sufficient to avoid this effect.}. If
this correction is made to each pulse, the artifact is removed
completely from the signal and all fluctuation statistics can
generally be interpreted correctly.

In some circumstances, extension of the approach to effect a change in
sampling interval (by dropping coefficients in the frequency domain)
might be indicated. This is often necessary in the analysis of
simultaneous multi-observatory data sets due to the employment of
different sampling intervals, and may also become important in future
high-resolution studies of close-orbit binary pulsars where the
apparent period can change significantly over the course of an
observation. We suggest the technique be investigated as a means for
avoiding the deleterious effects of time-domain re-sampling noticed by
\citet{khk+01}\footnote{Although we also note that for some purposes,
including application of the LRCCF between observatories as in
\citet{khk+01}, the synthesis of synchronous samples is unneccessary and
should be avoided.}.

In theory, the above procedure allows absolute alignment in longitude
of observations at different epochs and frequencies. In practice, the
published dispersion measure may derive from arbitrary profile
alignment procedures, and previously published timing ephemerides may
not extrapolate beyond the date range from which they derive
accurately enough for phase alignment between different epochs.  For
these reasons, we did not attempt absolute alignment, and our
longitude axes include an arbitrary offset.

\subsection{Accumulation of Statistics}
The first step of processing after forming the longitude/time array was
to correct for an absolute offset due to the  system noise power.
To reduce the effect of slow variations in the system temperature
we subtracted a running baseline, computed as the mean over the
surrounding $\pm$ 1000 pulses of all samples in a defined ``off-pulse''
longitude interval. Denoting the result as $S_{ij}$ where $i$ and $j$
are indices in pulse longitude and pulse number, we then computed the
(normalized) average pulse profile as

\begin{equation}
\mu_i = \frac{1}{N}\sum_{j=0}^{N-1} S_{ij},
\end{equation}
\noindent where $N$ is the number of pulses analyzed. For convenience,
at this point we re-normalized $S_{ij}$ so that the peak bin of
$\mu_i$ had a value of 1.  We then computed a version of the data from
which the average profile was subtracted:
\begin{equation}
S'_{ij} = S_{ij} - \mu_i.
\end{equation}

The remaining analysis involved the integration of three kinds of
correlation statistics that differ in their treatment of the longitude
dimension. These are the full two-dimensional autocorrelation function
(2DACF), the longitude-resolved autocorrelation function (LRACF) and
the longitude-resolved cross-correlation function (LRCCF).  

To examine the autocorrelation statistics of the signal while keeping
computational time and space requirements manageable, we divided the
input pulse series into $N/M$ $M$-pulse blocks of data and obtained
estimates of the fluctuation power spectra (LRFS and 2DFS) by
averaging the results of $N/M$ individual spectra computed from the
squared modulus of the appropriate DFTs. By application of the
Wiener-Khinchin theorem, the corresponding autocorrelation functions
were obtained via the DFTs of the power spectra. The choice of $M$
affected our sensitivity to nearly coherent modulation: since beyond
this length we summed {\it power} spectra, signals with decorrelation
time-scales longer than $M$ pulses would have been detected with
reduced significance. Signals with decorrelation timescales shorter
than $M$ could also have suffered since in that case the response would
be spread over several coefficients, however here the sensitivity can
be recovered by smoothing the power spectrum. Our approach was
therefore to choose the largest $M$ that was practical given the
computational and memory requirements.

In computation of the LRFS, each one-dimensional
spectrum was obtained using the squared modulus of the coefficients of
DFTs performed along a constant-longitude column of the longitude/time
array:
\begin{equation}
{\rm LRFS}_{ij}^{\rm raw}=\frac{1}{N}\sum_{k=0}^{N/M-1}\left|
                    \sum_{l=kM}^{(k+1)M-1}S'_{i,l}e^{-2\pi jl\sqrt{-1}/M}\right|^2.
\label{eq:LRFSobs}
\end{equation}
\noindent  To remove the effect of the variance
of the system noise, including non-white components due to
interference and system temperature variations not removed in the
pre-processing stages, we subtracted from each element of the spectrum
the corresponding value from the mean fluctuation spectrum of all
designated ``off-pulse'' longitude bins, dropping the superscript of
Eq. \ref{eq:LRFSobs} to denote the noise-corrected LRFS.

 We then performed DFTs along constant-longitude columns of the LRFS to
obtain the LRACF.  To avoid performing a cyclic convolution, the
individual LRFS were actually computed over $2M$ points, of which the
elements of one contiguous half were set to zero. The resulting LRFS
elements are not independent, so to facilitate proper evaluation of
the significance of spectral features, after using it to form the
LRACF we discarded the odd-numbered coefficients of the LRFS (which
has the same result as an $M$-point transform of the non-padded data).
To correct the LRACF result for edge normalization effects we divided
the result by $1-j/M$, where is $j$ the lag number. The result is:
\begin{equation}
{\rm LRACF}_{ij} =   \frac{1}{N(1 -j/M)} 
 \sum_{k=0}^{N/M-1}\sum_{l=kM}^{(k+1)M-1-j} S'_{il} S'_{i(l+j)}.
\end{equation}
\noindent The zero lag of the LRACF gives the longitude-resolved
variance:
\begin{eqnarray}
\sigma^2_i &=& {\rm LRACF}_{i0} \\
            &=& \frac{1}{N}\sum_{j=0}^{N-1}(S'_{ij})^2.
\label{eq:sigmasq}
\end{eqnarray}
\noindent From this we computed the modulation index as $m_i =
\sigma_i / \mu_i$. Since the zero lag of the LRACF is simply the sum
of all elements of the corresponding fluctuation power spectrum, we
may view the variance in a given longitude bin as having contributions
from different frequencies (Parseval's theorem). We made use of this
fact to compute modulation indices with the exclusion of the lowest
frequency bin, since it often contained a strong component which, by
virtue of its low frequency and longitude-independent modulation
index, we attribute to interstellar scintillation.

The 2DFS was derived through a similar division of the data sequence
into $M$-pulse blocks. Since it integrates in longitude, we only
include a designated on-pulse longitude interval in the analysis.
The squared modulus of the two-dimensional DFT
of such blocks gives the 2DF power spectrum, which was integrated over
all blocks:
\begin{eqnarray}
{\rm 2DFS}_{ij}^{\rm raw}=& & \frac{1}{NB}\sum_{k=0}^{N/M-1} \nonumber\\
 & &   \left| \sum_{l=kM}^{(k+1)M-1}\sum_{b=0}^{B-1}
     S'_{i,l}e^{-2\pi \sqrt{-1}(ib/B+jl/M)}\right|^2,
\end{eqnarray}
\noindent where $B$ is the number of longitude bins.  When possible,
to correct for the autocorrelation characteristics of the noise and
interference, the raw 2DFS of an equal-sized ``off-pulse'' longitude
interval (or the mean from several intervals) was computed and
subtracted from the raw on-pulse spectrum.  

The 2DFS has two dimensions that are both dimensionless
frequencies. In this work we refer to quantities in the pulse number-
and longitude-associated axes as temporal and longitudinal frequencies
respectively, and plot the spectra with the longitudinal frequency
axis as the abscissa. We usually use units of (subpulse) cycles per
(spin) period (cpp) for these quantities, which in the case of
longitudinal frequencies also corresponds to degrees of subpulse phase
per degree of longitude.  The 2DFS is periodic (due to aliasing with
the finite sample interval) and symmetric about the diagonal (due to
the real-valued input), obeying the relation ${\rm 2DFS}_{ij} = {\rm
2DFS}_{(\pm i+cB)(\mp j+dM)}$ where $c$ and $d$ are arbitrary integers,
and can therefore be fully specified by computing a contiguous region
of either $B/2+1 \times M$ coefficients, or $B \times M/2+1$
coefficients.  In this work we plot the region given by $i \in
[-B/2:B/2)$ (i.e. longitudinal frequencies between
$-180\degr/\Delta\phi$ and $180\degr/\Delta\phi$, where $\Delta\phi$
is the longitude sampling interval) and $j \in [0:M/2]$ (i.e. temporal
frequencies between 0 and 0.5 cpp), which generally avoids having
responses ``wrap'' across borders in the planar projection, and
directly gives information concerning the apparent sense and rate of
any subpulse drift pattern.

The two-dimensional DFT of the 2DFS gives a form of Two-Dimensional
Autocorrelation Function (2DACF), a function which, like the LRFS, has
seen frequent use in the past for drifting subpulse analysis
(e.g. \citealt{tmh75}). Again, it is important to pad the data with
zeroes (in both axes) to avoid computing a cyclic convolution, and to
discard odd-numbered rows and columns of the resultant 2DFS to avoid
misinterpereting the significance of spectral features. The correction
for edge normalization effects was performed through multiplication by
$1-j/M$; since the pulsar signal is not expected to extend beyond the
longitude boundaries, additional normalization by $1-i/B$ is not
appropriate. The result is:
\begin{eqnarray}
{\rm 2DACF}_{ij} = &&  \frac{1}{NB(1 -j/M)} \nonumber\\
 &&\sum_{k=0}^{N/M-1}\sum_{l=kM}^{(k+1)M-1-j} \sum_{b=0}^{B-1-i} 
    S'_{bl} S'_{(b+i)(l+j)}.
\end{eqnarray}
\noindent The zero pulse-number lag of the 2DACF gives the
single-pulse ACF used by \citet{jap01}:
\begin{eqnarray}
{\rm ACF}_i &=& {\rm 2DACF}_{i0} \\
            &=& \frac{1}{N}\sum_{j=0}^{N-1} 
                 \sum_{b=0}^{B-1}S'_{bj} S'_{(b+i)j} .
\end{eqnarray}

As an additional diagnostic of pulse-to-pulse correlations,
we computed the LRCCF, which measures the covariance between every
pair of longitude bins as a function of pulse number offset. We
define the LRCCF as
\begin{equation}
{\rm LRCCF}_{i_1i_2k} = \frac{1}{N-k}\sum_{j=0}^{N-k-1}S'_{i_1j}S'_{i_2(j+k)}.
\label{eq:lrccf}
\end{equation}
\noindent Our definition differs somewhat from that of \citet{pop86},
who normalized by $1/\sigma_{i_1}\sigma_{i_2}$ to obtain a true
correlation rather than a covariance. We find the covariance
preferable in the evaluation of the significance of low S/N results
since the noise levels are identical in all bins of each
two-dimensional ($i_1, i_2$) map.  The computation of the LRCCF could,
in principle, be incorporated within the above scheme by integrating
the conjugate product of complex LRFS pairs and taking the DFT of the
result, however the computational complexity is much greater than for
the autocorrelation techniques above.  Since we were only interested
in the first $\sim 10$ ($\ll M$) lags we found it more efficient to
compute directly from Eq.~\ref{eq:lrccf}. We note however that the
frequency domain counterpart (the Longitude-Resolved Fluctuation
Cross-Power Spectrum) could be useful in simultaneous multi-frequency
observations for examining the presence, strength and relative phase
of periodic modulations between different longitudes and frequencies.

Finally, although individual pulse detections were not the focus of
this work, the assembled data set provides an opportunity for a survey
of giant pulse behaviour in recycled pulsars. For this reason, we
searched each observation for significant single pulses by convolving
each pulse with boxcars of widths of 1, 2, 3, 4, 6, 8, 12 \ldots
$W_{\max}$ bins (where $W_{\max}$ is approximately the width of the
on-pulse region of the average profile) and selecting those pulses
containing convolved samples with a single-trial significance
exceeding $6\sigma$. We then computed the energy of the selected
pulses as the sum of intensity values over an interval centered on the
same longitude as the detection boxcar but with twice the width, and
expressed these as a fraction of the energy of the average profile.

\section{Results}
\subsection{Non-Detections}
\label{sec:nondet}
In the observations of PSR J0218+4232, J0034$-$0534, J0613$-$0200,
J1643$-$1224 and B1937+21, with the exception of very low frequency
fluctuation attributable to scintillation, we found no detectable
modulation in any of the measured statistics. As with all pulsars
observed, this included searches of the LRFS for transform lengths
from $2^9$ to $2^{17}$ pulses (unless the number of pulses observed
was less than this), from which one may derive limits on the presence
of quasi-periodic modulation. Considering a longitude-resolved
variance computed over a given interval $\Delta\nu$, the noise terms
in the sum result in an approximately Gaussian distribution (central
limit theorem) with a standard deviation proportional to
$\Delta\nu^{1/2}$. For a given noise level estimated from the
off-pulse parts of a LRFS, we take the square root of the 3-$\sigma$
point of the noise in a given estimated variance
(Eq. \ref{eq:sigmasq}) as the minimum detectable standard
deviation. These values are given for each observation in Table
\ref{tab:obs}.  The minimum detectable standard deviation for a given
frequency integration is given by $\sigma_g(\Delta\nu/{\rm
cpp})^{1/4}$, which we have tabulated as $\sigma_{\rm m}$ for the full
modulation index ($\Delta\nu = 0.5$ cpp) and for unresolved
(i.e. nearly coherent) features in the longest transform, as
$\sigma_{\rm c}$ ($\Delta\nu = 1/M_{\rm max}$ cpp, $M_{\rm
max}=2^{17}$ unless noted). {\changed With the exception of PSR B1937+21, the
four pulsars for which we have not detected modulation are those with
the worst detection limits, and for this reason the results cannot be
taken as evidence for low modulation indices.}

The search for bright individual pulses was successful in several
pulsars (noted below), however with the exception of PSR B1937+21, no
pulsar produced clearly detectable {\em giant} pulses in our
observations.  For PSR J0613$-$0200, one pulse barely exceeded the
threshold, occurring within the (broad) on-pulse region, with an
energy exceeding the mean by a factor of 20 and a width of one
bin. Further studies should be made before attributing this detection
to a giant pulse phenomemon.  For all pulsars, the $6$-$\sigma$ energy
limit for individual samples is presented in Table \ref{tab:obs} as
$E_{\rm min}$; for wider pulses the limit scales as $n^{1/2}$ where
$n$ is the number of samples to add.

\subsection{PSR J1012+5307}
\begin{figure*}
\resizebox{0.95\hsize}{!}{\includegraphics{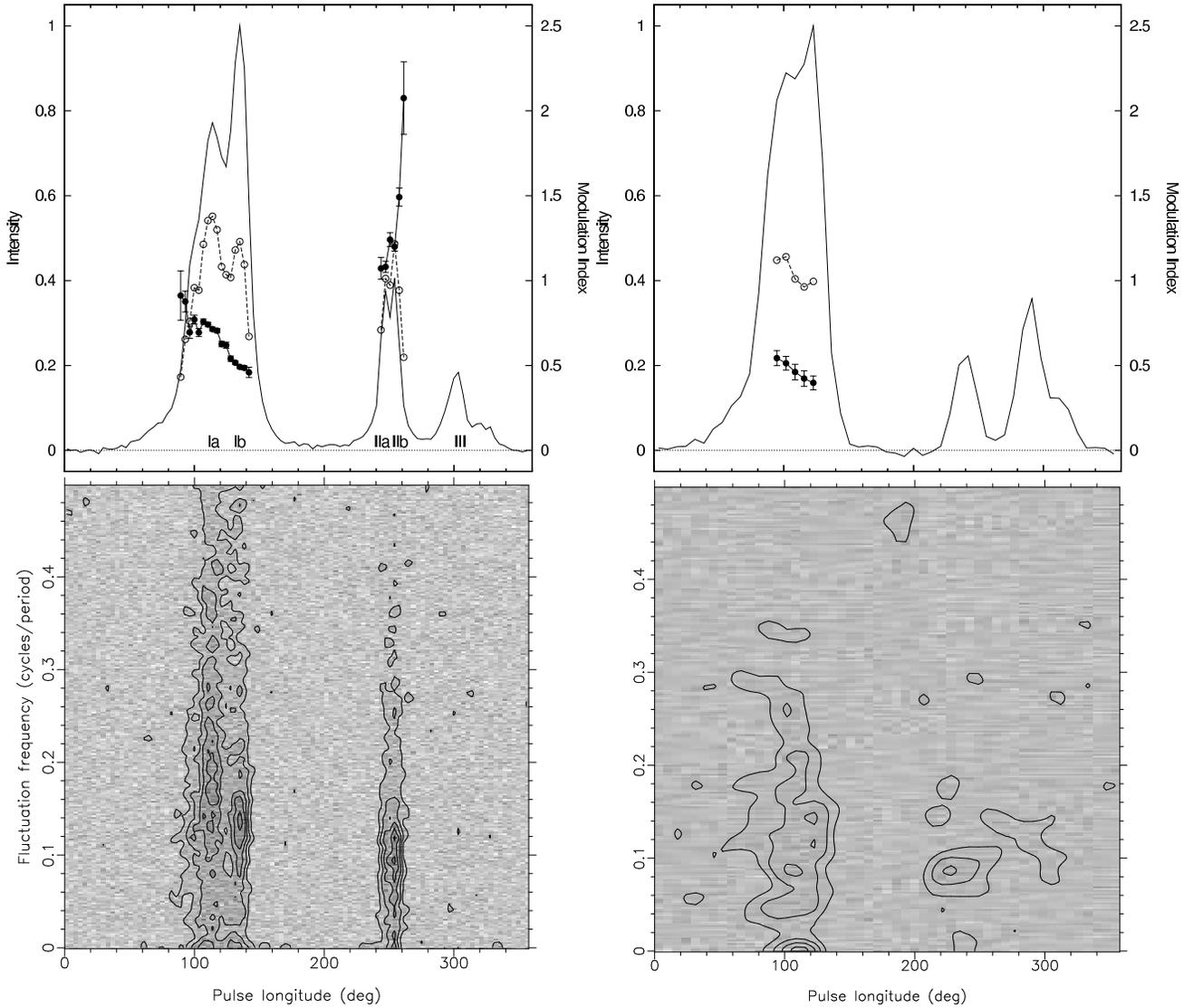}}
\caption{Longitude-resolved modulation statistics for PSR J1012+5307
at 1380 MHz (left) and 840 MHz (right).  Top panel: mean pulse profile
(top, solid lines), standard deviation (unfilled circles) and
modulation index (filled circles, right-side axis scales) for PSR
J1012+5307 at 1380 MHz (left) and 840 MHz (right). The names used for
components are indicated in the top left panel. The standard deviation
and modulation index are subject to a 3-$\sigma$ threshold in the
power values from which they are derived, to ensure their
significance. A strong component in the DC frequency bin of the LRFS
is attributed to scintillation and was not included in the calculation
of these statistics. Bottom panel: LRFS, shown as a greyscale computed
with $M=512$, along with contours of Gaussian-smoothed version of the
same spectrum of FWHM $0\degr \times 0.02$ cpp (1380 MHz) and $20\degr
\times 0.02$ cpp (840 MHz). }
\label{fig:1012lr}
\end{figure*}

\begin{figure}
\resizebox{0.95\hsize}{!}{\includegraphics{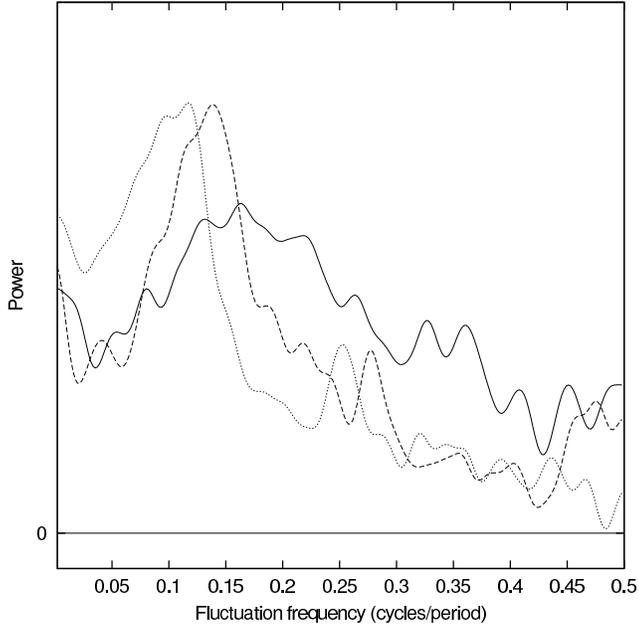}}
\caption{Fluctuation spectra for PSR J1012+5307 at 1380 MHz, from peak
bins in components Ia (solid), Ib (dashed) and IIb (dotted; IIa
appeared similar and is omitted for clarity). The raw spectra were
convolved with a Gaussian of FWHM 0.02 cpp before plotting.}
\label{fig:1012lrfs}
\end{figure}

The fluctuation spectra for PSR J1012+5307 (Fig. \ref{fig:1012lr})
indicate the presence of quasi-periodic modulation in all three major
profile components. Moreover, it is apparent in the 1380 MHz
observation that the peak frequency of the modulation varies with
longitude, showing different values in components Ia, Ib and II. To
better examine the differences between the components, we have plotted
fluctuation spectra from components Ia, Ib and IIb in
Fig. \ref{fig:1012lrfs}.  From this figure it is apparent that
although the overall depth of modulation in component Ia is higher
than than that in Ib, its quasi-periodic feature is broader in
frequency space. We also note the presence of narrow features in the
spectra of components Ib and IIb at frequencies close to twice the
frequency of peak power. The significance of these features deteriorates
with a narrower smoothing function, indicating that their intrinsic width
is close to that seen in Fig. \ref{fig:1012lrfs}.

We detected 70 individual bright pulses in the 1380~MHz observation,
the strongest of which had an energy five times the mean pulse energy
for this obervation. The pulse profile formed by adding these pulses
shows relatively more emission in components Ia and IIb compared to Ib
and IIa, consistent with the higher modulation indices found in these
regions.

\begin{figure}
\resizebox{0.95\hsize}{!}{\includegraphics{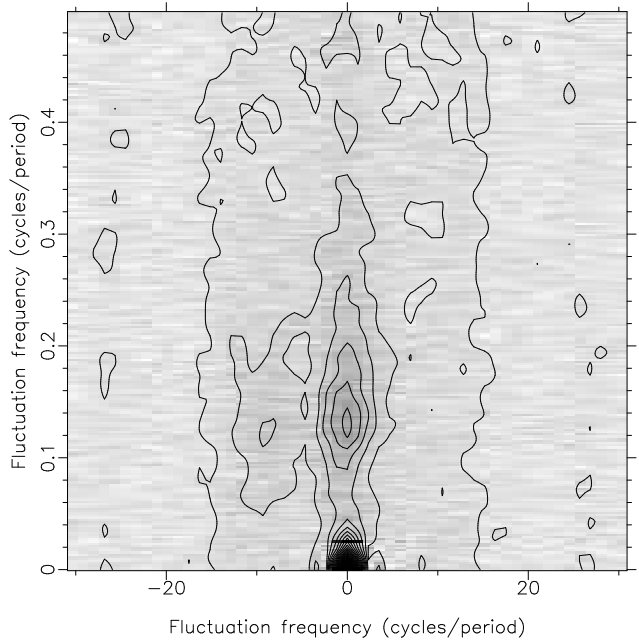}}
\caption{2DFS for component I of PSR J1012+5307 at 1380 MHz, using
$M=512$ (greyscale) and smoothing by convolution with a Gaussian of
FWHM $0\times0.02$ cpp (contours).}
\label{fig:10122dfs}
\end{figure}

The two-dimensional fluctuation spectra were contaminated by the
presence of broad stripes aligned with the temporal frequency axis,
which we were unable to subtract due to the lack of sufficiently
wide off-pulse longitude intervals. Nevertheless, it is clear that
the modulation power in the component I is divided between two main
regions, one with zero frequency in the longitude axis, and one with a
freqency of $\sim -10$ cpp (Fig. \ref{fig:10122dfs}).  The latter
component indicates the presence of quasi-periodic drifting subpulses,
with a longitude separation of $\sim 40\degr$ between subpulses. Since
the component I is double-peaked, one might also consider the
alternative  of longitude-stationary subpulse modulation with a
simple phase offset between the two halves of the profile
component. However, this can be rejected because the peak frequency
obtained could not be more than $180\degr$ divided by the component
separation of $\sim 40\degr$, i.e.  $\sim 5$ cpp, where there is
certainly no peak in the spectrum.  The spectral component centered
around zero longitudinal frequency indicates the presence of
longitude-stationary quasi-periodic modulation, while its relatively
narrow extent suggests it extends across most of the component I.

It is  conceivable that the two spectral components in the 2DFS
are in fact associated with with same modulation feature, if the
apparent temporal frequency offset  is actually
spurious. In this case the shifted, double-peaked nature of the
spectrum in the longitudinal axis is indicative of linear
drifting with a frequency of $\sim 5$ cpp and a phase step
between opposite halves of the profile component \citep{esv03}.
\begin{figure}
\begin{tabular}{c}
\resizebox{0.8\hsize}{!}{\includegraphics{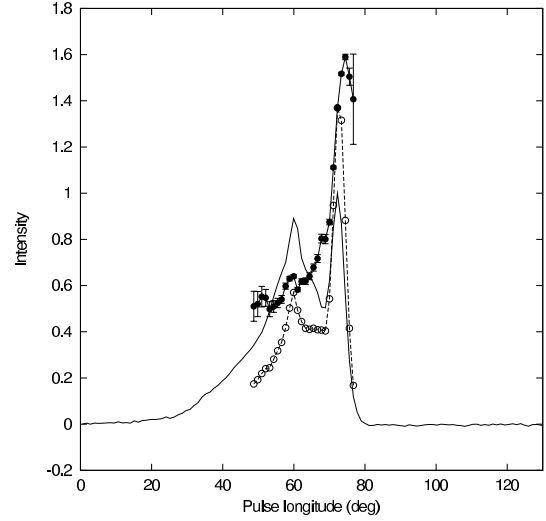}} \\
\resizebox{0.8\hsize}{!}{\includegraphics{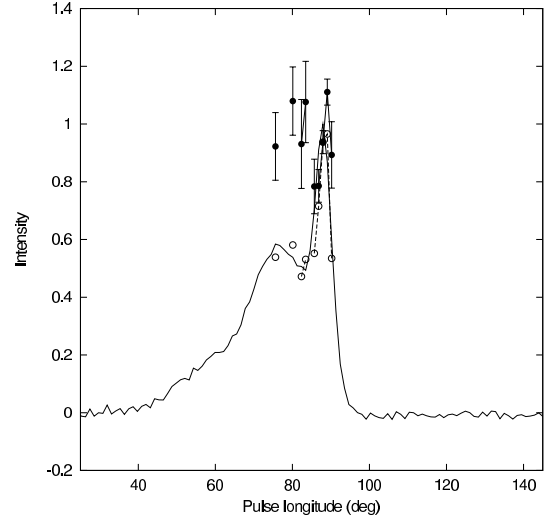}} \\
\resizebox{0.8\hsize}{!}{\includegraphics{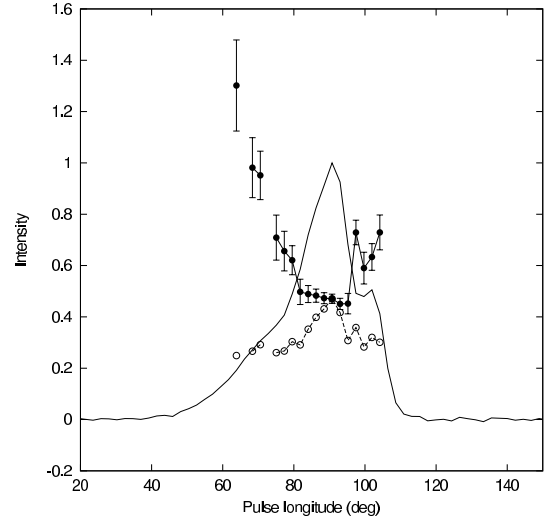}} 
\end{tabular}
\caption{Average profile, standard deviation and modulation index for
PSR J1022+1001 at 1380 (top), 840 (middle) and 328 MHz (bottom).  See
caption for Fig. \ref{fig:1012lr}, but here the modulation index is
plotted with the same scale as the other quantities.}
\label{fig:1022lr}
\end{figure}

\begin{figure}
\resizebox{\hsize}{!}{\includegraphics{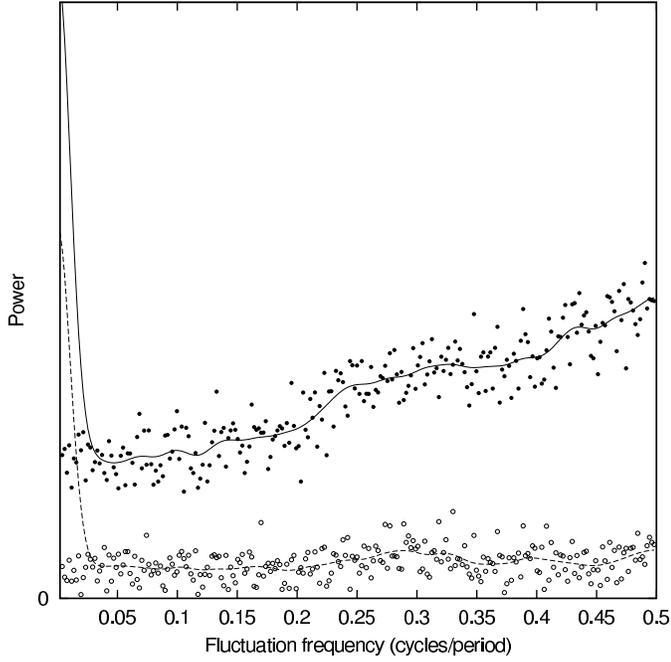}} 
\caption{Fluctuation spectra from peak longitude bins of the leading
(open circles, dashed line) and trailing (filled circles, solid line)
components of PSR J1022+1001. The points show the measured spectrum
($M=512$) in the peak bin, while the solid and broken lines show the
spectra from the peak and surrounding bins respectively, after
smoothing by convolution with a Gaussian of FWHM 0.02 cpp. The
low-frequency excess is due to very low freqency modulation
(attributed to scintillation), broadened by the Gaussian smoothing
kernel.}
\label{fig:1022lrfs}
\end{figure}

\begin{figure}
\centering{
\resizebox{0.86\hsize}{!}{\includegraphics{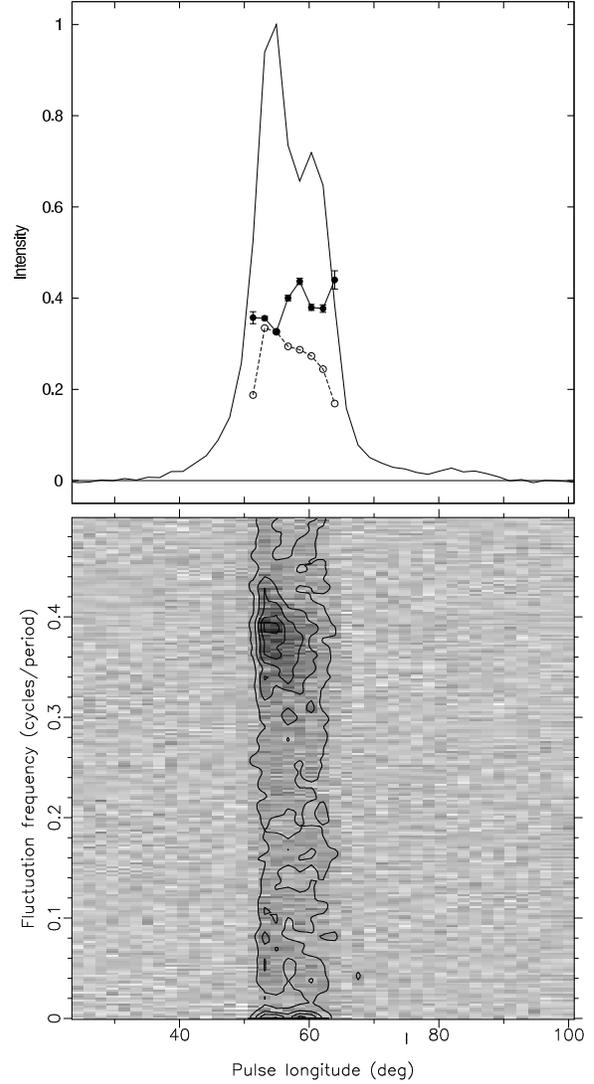}}
\caption{Longitude-resolved statistics for PSR J1518+4904. See caption
for Fig.~\ref{fig:1012lr}.  }
\label{fig:1518lr}}
\end{figure}

\begin{figure}
\resizebox{\hsize}{!}{\includegraphics{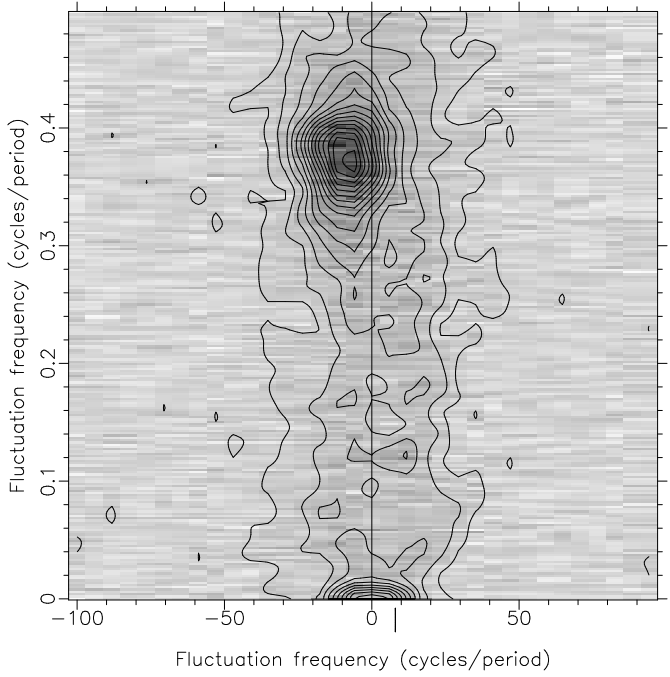}}
\caption{2DFS for PSR J1518+4904 at 1355 MHz. See caption for 
Fig. \ref{fig:10122dfs}.}
\label{fig:15182dfs}
\end{figure}
\begin{figure}
\resizebox{\hsize}{!}{\includegraphics{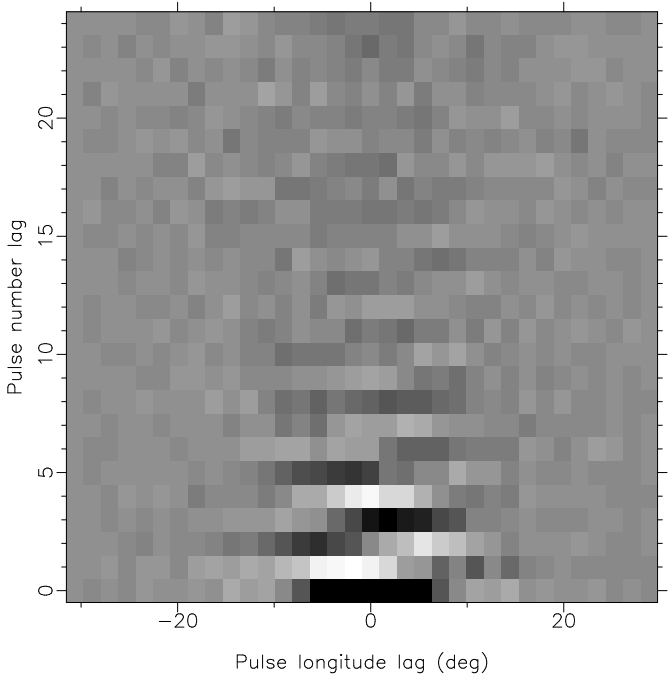}}
\caption{Lags 0--24 of 2DACF for PSR J1518+4904 at 1355 MHz. Black and
white values correspond to the most positive and negative values
respectively, outside the zero lag.}
\label{fig:15182dacf}
\end{figure}
\begin{figure}
\resizebox{\hsize}{!}{\includegraphics{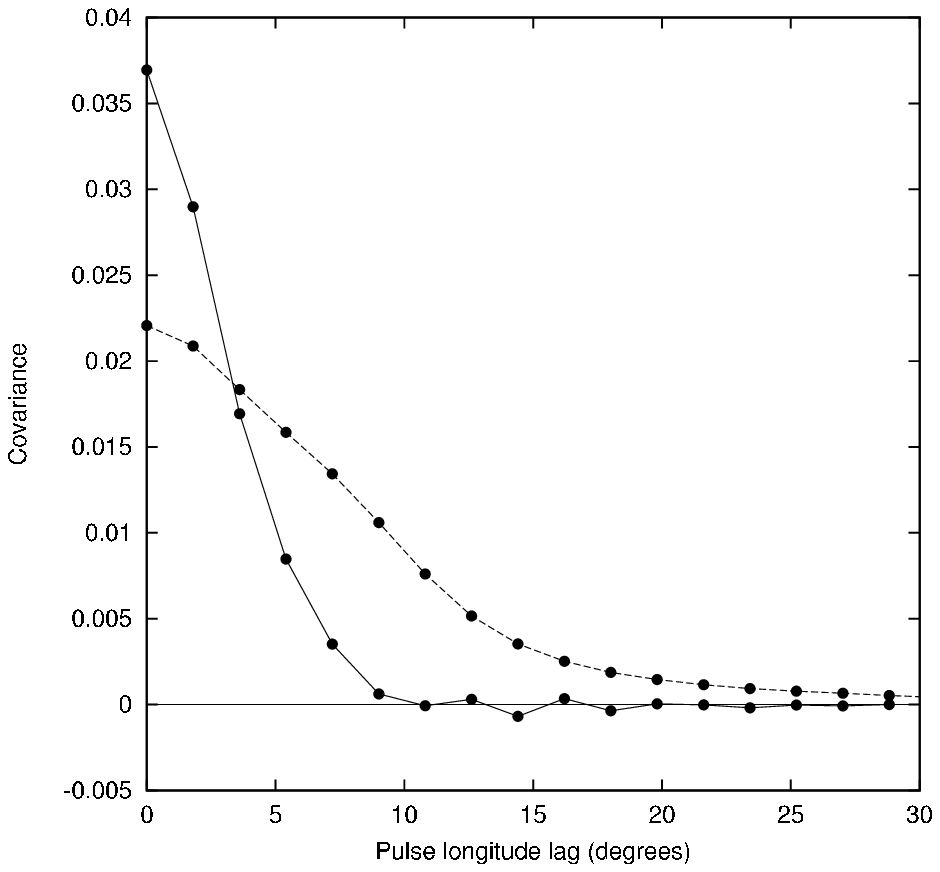}}
\caption{Single-pulse ACF (solid line) and ACF of the average profile
(dashed line) for PSR J1518+4904 at 1355 MHz.}
\label{fig:1518acf}
\end{figure}

\subsection{PSR J1022+1001}
This pulsar showed non-periodic modulation at 328, 840 and 1380~MHz
(Fig. \ref{fig:1022lr}).  At 1380~MHz, the trailing component is seen
to be very strongly modulated, a fact corroborated by the dominance of
this component in the average profile derived from the $\sim 10^4$
individual pulses we detected. In this observation, the brightest
pulses had an energy $\sim 6.5$ times the mean, while the distribution
of energy measured in a longitude interval spanning the trailing
component was consistent with an exponential or power-law.

The fluctuation spectra of the bins near the two peaks of the average
profile at 1380~MHz are shown in Fig. \ref{fig:1022lrfs}. While the
leading component shows a flat fluctuation spectrum, the trailing
component proceeds smoothly to a maximum at the Nyquist
frequency. This corresponds in the autocorrelation function to a
strong negative spike at a lag of one pulse. There is no evidence for
any periodicity associated with this correlation.

\subsection{PSR J1518+4904}
\begin{figure*}
\resizebox{0.9\hsize}{!}{\includegraphics{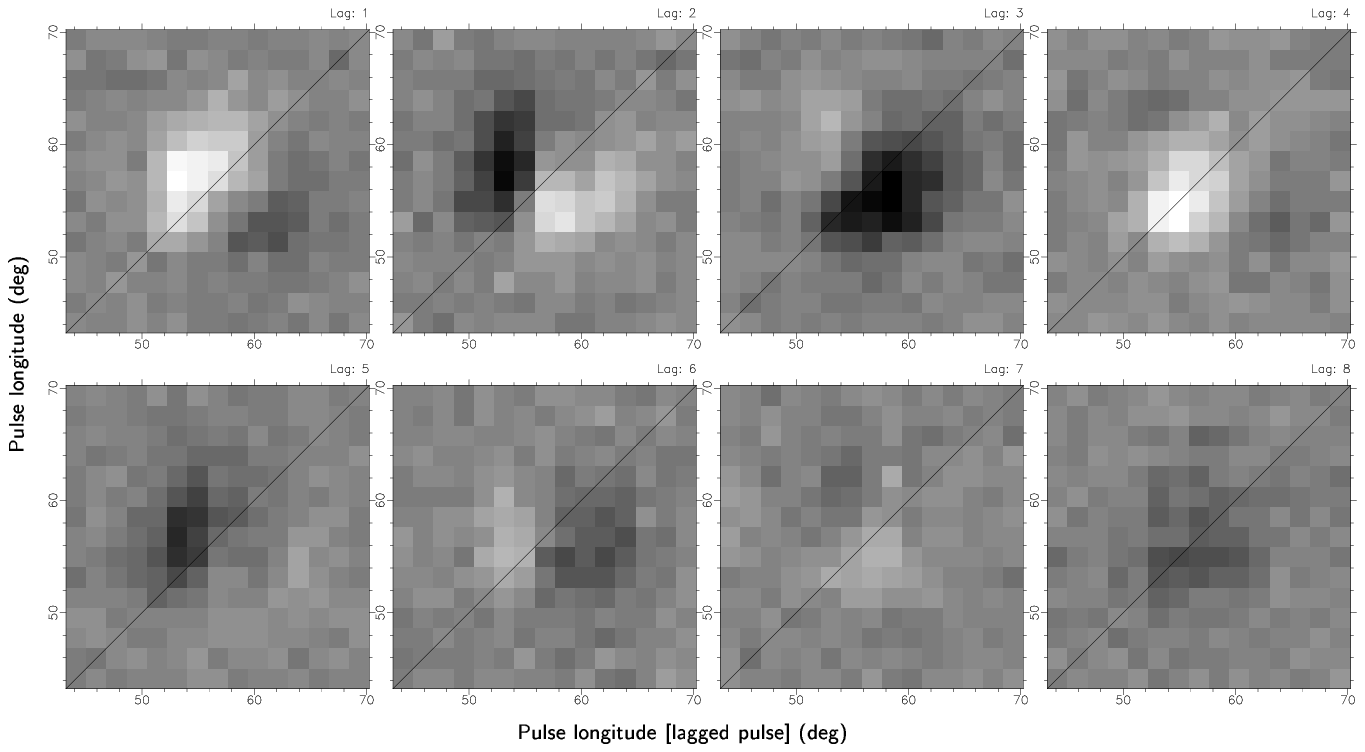}}
\caption{Lags 1--8 of LRCCF for PSR J1518+4904 at 1355 MHz. Grey levels
are set as in Fig.\ref{fig:15182dacf}. The solid lines connect bins
with the same longitude value in the lagged and non-lagged pulse
(i.e. the autocorrelations).}
\label{fig:1518lrcc}
\end{figure*}

We detected quasi-periodic modulation throughout the double-peaked main
profile component of PSR J1518+4904 (Fig. \ref{fig:1518lr}), with a
frequency of approximately 0.38 cpp. In the peak bin this modulation
contributes about half of the total modulation power, however this
fraction decreases significantly toward the trailing part of the
profile.  The 2DFS (Fig. \ref{fig:15182dfs}) shows a corresponding
component with its peak centered at a longitudinal frequency $\sim
-10$ cpp, corresponding to a phase slope of about half a cycle across
the pulse window. The 2DACF (Fig. \ref{fig:15182dacf}) confirms this
description, showing the characteristic repeating pattern of drifting
subpulses. The fact that for any given pulse number lag the 2DACF is
singly-peaked confirms what one might expect with the relatively small
phase slope, namely that usually only one subpulse is present as part
of the quasi-periodic pattern. Fig.  \ref{fig:1518acf} shows this more
clearly for the zero lag, and confirms that the subpulses are
significantly narrower than the average profile. The drift pattern is
also manifest in the LRCCF (Fig. \ref{fig:1518lrcc}), from which it is
clear that drifting of the modulation applies over the entire
longitude window defined by the average profile.

We detected 1650 individual pulses from this pulsar, the
brightest of which had an energy 4.8 times the mean.

\subsection{PSR J1713+0747}
\begin{figure*}
\begin{tabular}{cc}
(a)\resizebox{0.35\hsize}{!}{\includegraphics{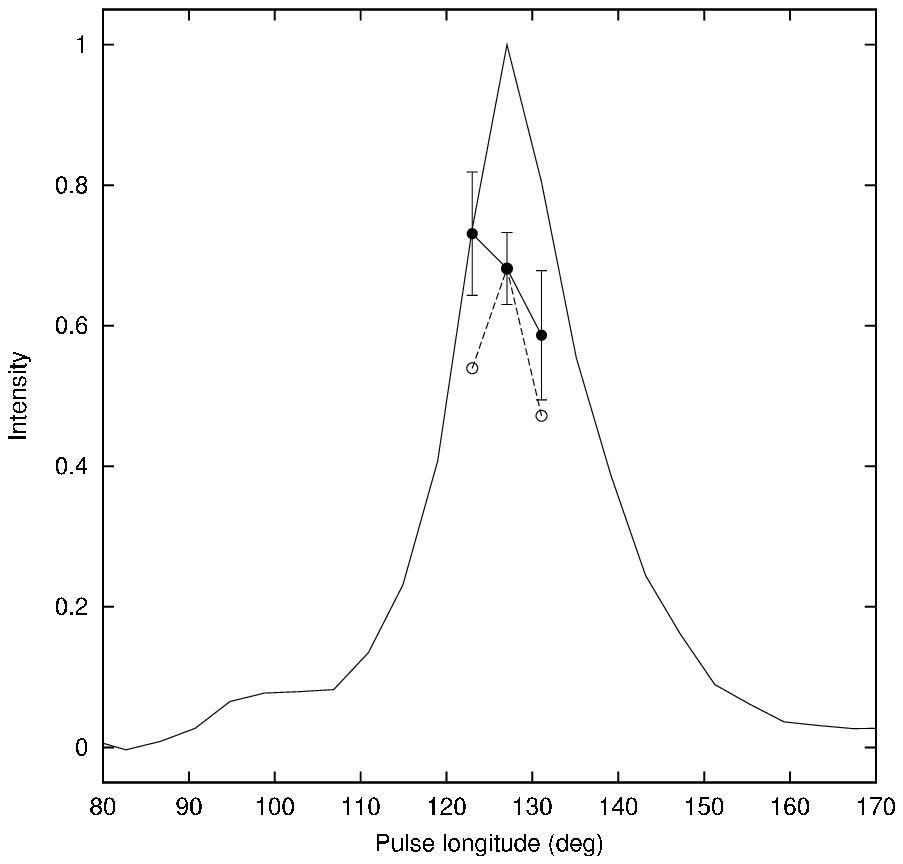}}
&
\resizebox{0.35\hsize}{!}{\includegraphics{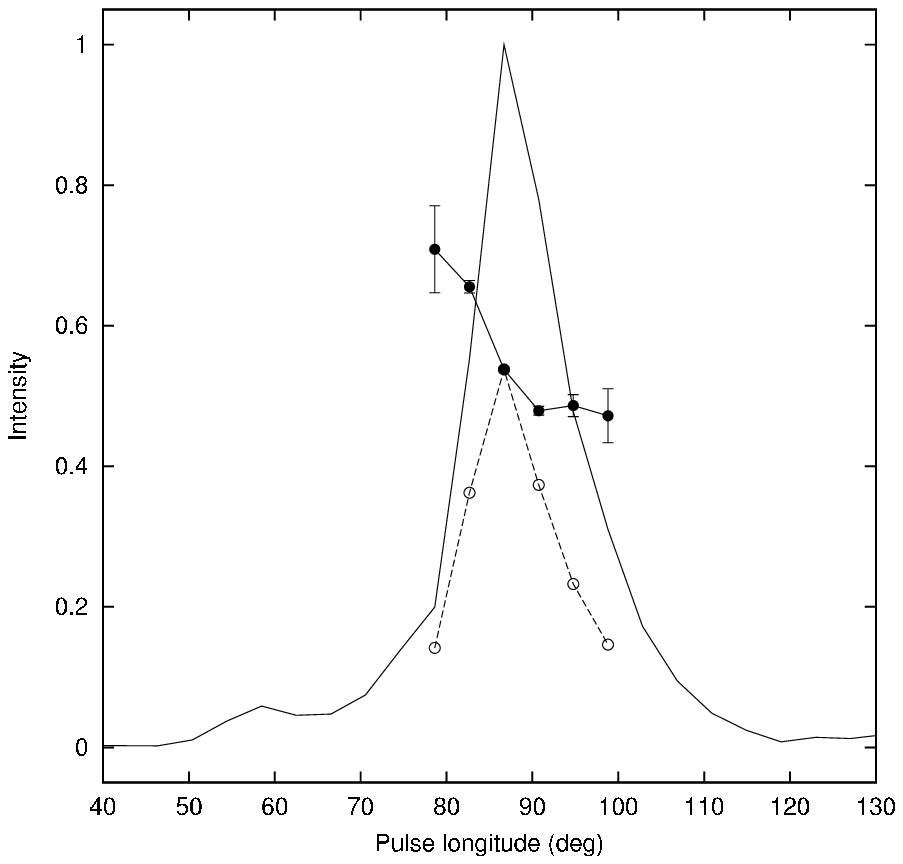}} (b)\\
(c)\resizebox{0.35\hsize}{!}{\includegraphics{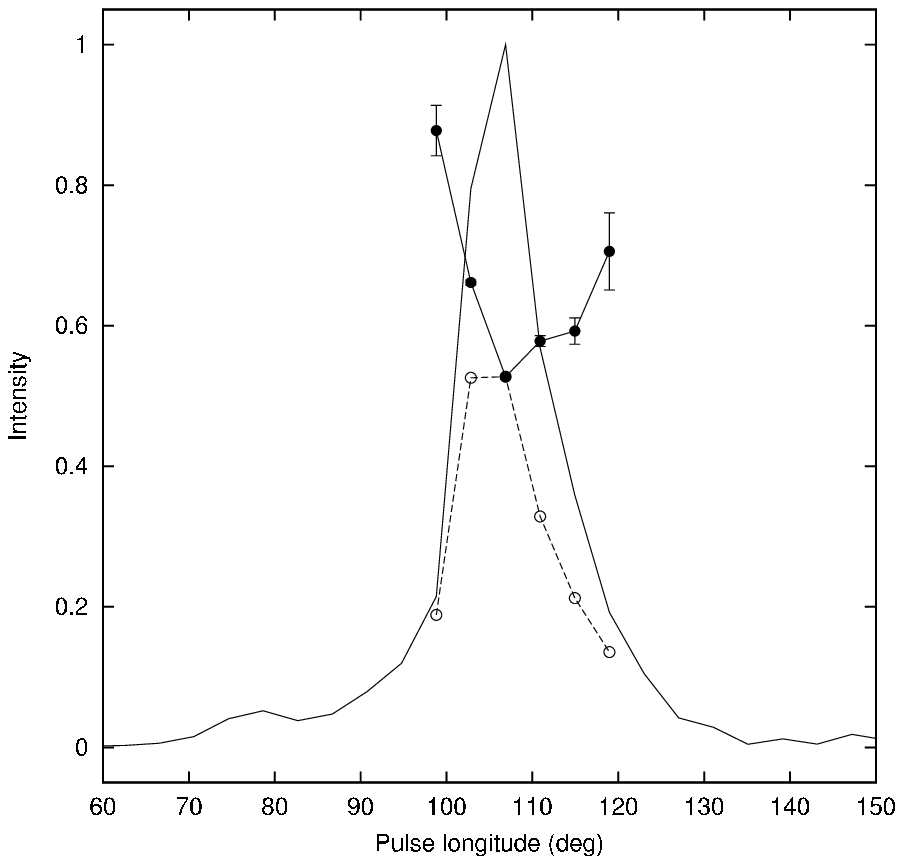}} &
\resizebox{0.35\hsize}{!}{\includegraphics{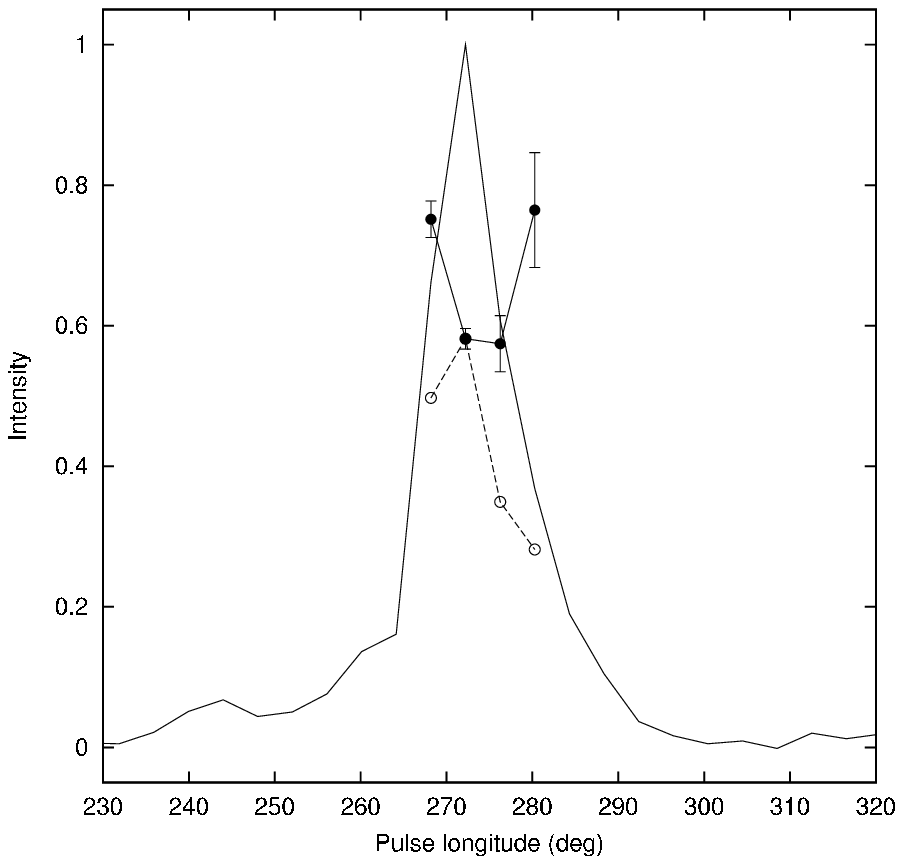}} (d)\\
\end{tabular}
\caption{Average profile, standard deviation and modulation index for
 PSR J1713+0747 at (a) 840 MHz, (b) 1190 MHz, (c) 1700 MHz and (d)
 2240 MHz. See also caption for Fig. \ref{fig:1012lr}.}
\label{fig:1713modind}
\end{figure*}

\begin{figure*}
\begin{tabular}{cc}
\resizebox{0.5\hsize}{!}{\includegraphics{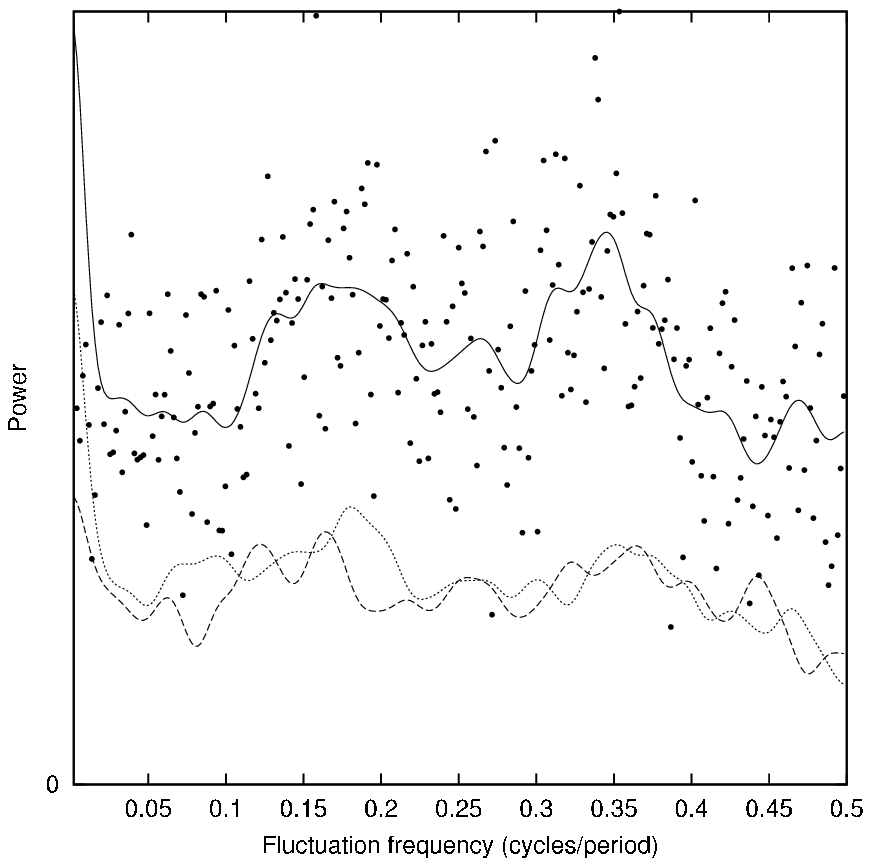}} &
\resizebox{0.5\hsize}{!}{\includegraphics{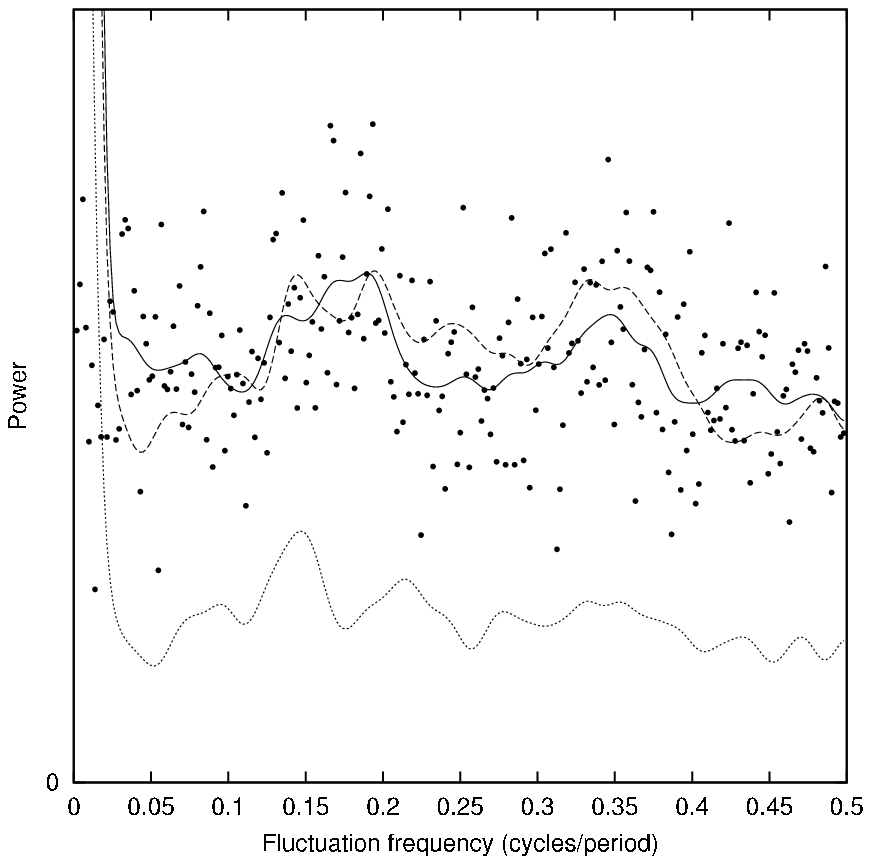}}
\end{tabular}
\caption{Fluctuation spectra from significant bins of the LRFS, at
1190 MHz (left) and 1700 MHz (right), for PSR J1713+0747. See also
caption for Fig. \ref{fig:1022lrfs}. }
\label{fig:1713lrfs}
\end{figure*}

We have detected modulation in observations at PSR J1713+0747 from 840
to 2240 MHz (Fig. \ref{fig:1713modind}). The evolution in the pulse
profile and modulation index is very minor over this range, however to
aid in the assesment of the longitude-dependence of the modulation
index under the low longitude resolution available, we have provided
plots for all four frequencies. It is clear that the leading edge
shows a higher modulation index than other parts of the profile. A
selection of one-dimensional fluctuation spectra are shown in
Fig. \ref{fig:1713lrfs}. Two broad maxima at frequencies of $\sim
0.17$ and $\sim 0.35$ cpp are clearly seen.

We detected $\sim 50$ individual pulses at 1190~MHz, with energies up
to 6 times the mean.

\subsection{PSR J1918$-$0642}
\begin{figure}
\resizebox{\hsize}{!}{\includegraphics{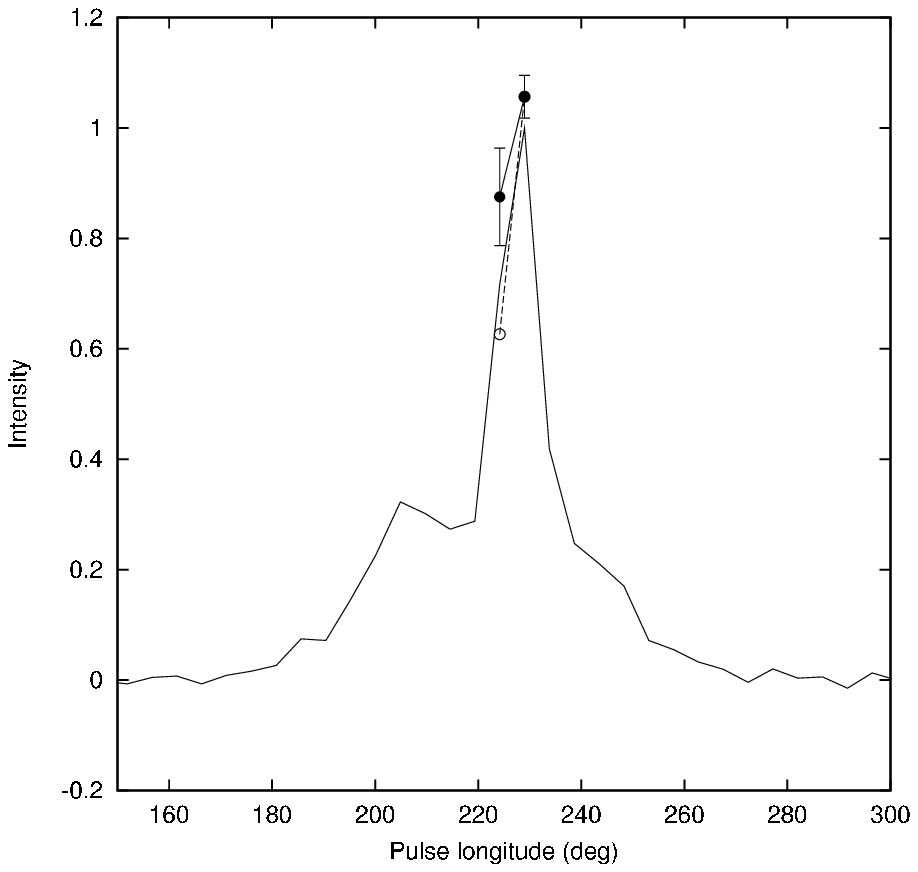}}
\caption{Average profile, standard deviation and modulation index for
PSR J1918$-$0642 at 1380 MHz. 
See caption for Fig. \ref{fig:1012lr}.}
\label{fig:1918modind}
\end{figure}

\begin{figure}
\resizebox{\hsize}{!}{\includegraphics{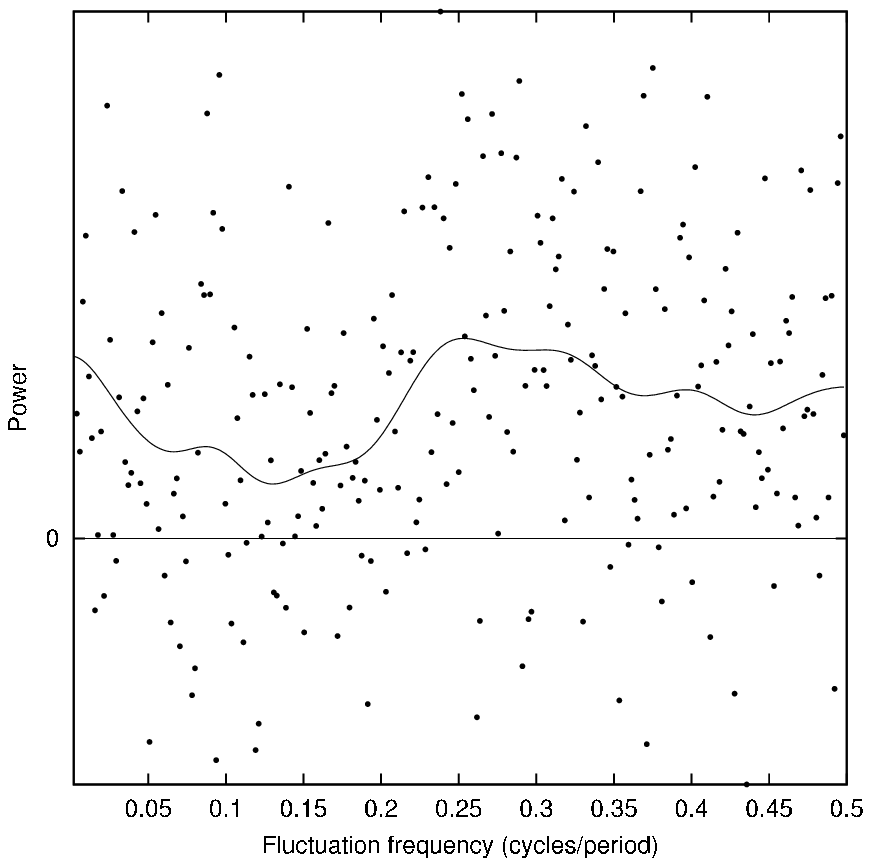}}
\caption{Fluctuation spectrum from the peak longitude bin of PSR J1918$-$0642,
at 1380 MHz. A Gaussian of FWHM 0.05 cpp was used to produced the smoothed
spectrum. (See caption for Fig. \ref{fig:1022lrfs}.)}
\label{fig:1918lrfs}
\end{figure}

We detected modulation in the strongest two longitude bins in an
observation of PSR J1918$-$0642 at 1380 MHz
(Fig. \ref{fig:1918modind}). There is some evidence for a
concentration of power between 0.25 -- 0.5 cpp in the fluctuation
spectrum (Fig. \ref{fig:1918lrfs}), however the limited sensitivity
makes this result uncertain. 


\subsection{PSR J2145$-$0750}
\begin{figure*}
\begin{tabular}{cc}
\resizebox{0.48\hsize}{!}{\includegraphics{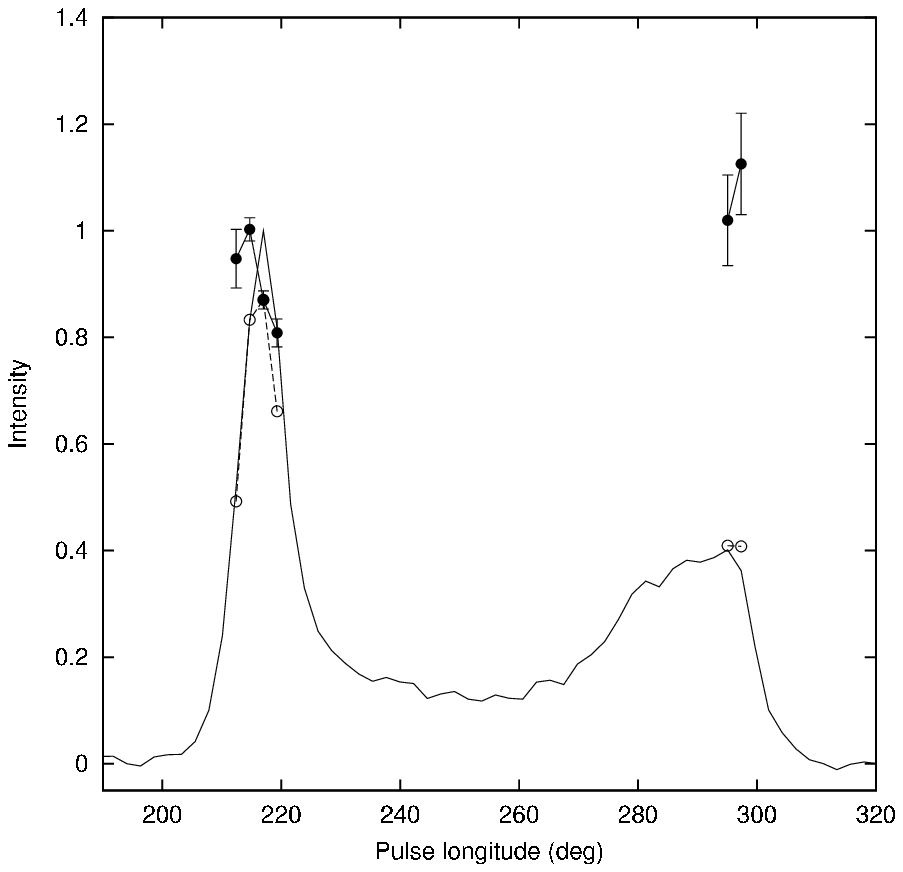}} &
\resizebox{0.48\hsize}{!}{\includegraphics{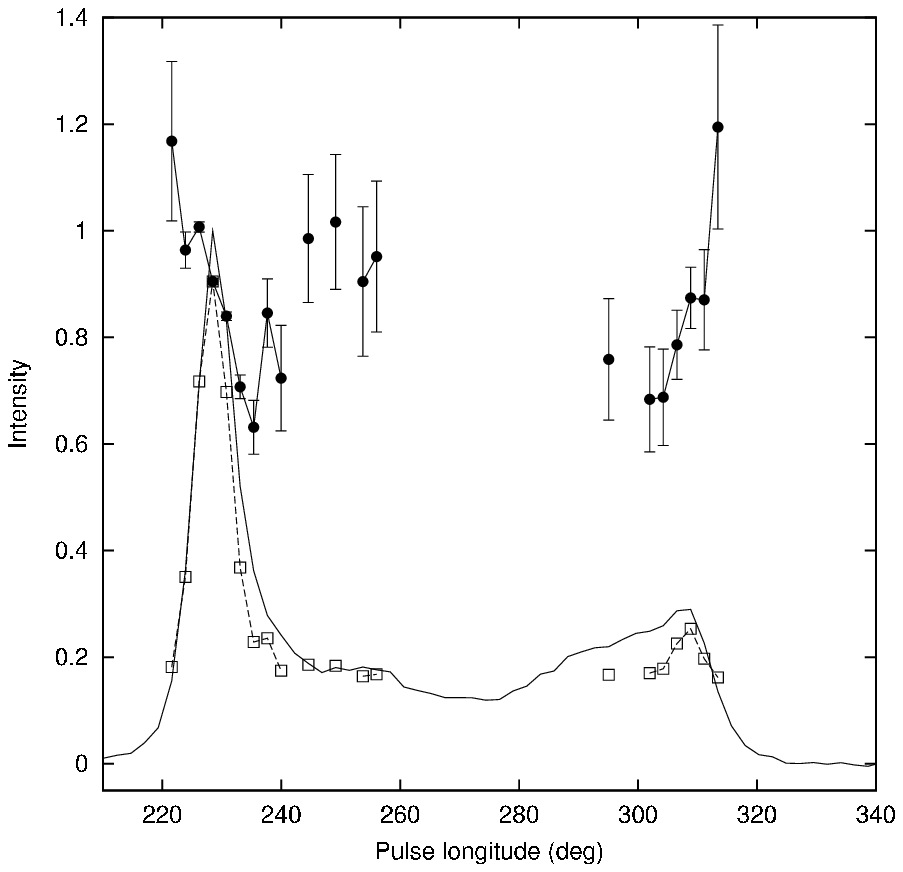}}
\end{tabular}
\caption{Longitude-dependent statistics for PSR J2145$-$0750 at
860 MHz (left) and 1380 MHz (right). See caption for Fig. \ref{fig:1012lr}.}
\label{fig:2145modind}
\end{figure*}

\begin{figure}
\resizebox{\hsize}{!}{\includegraphics{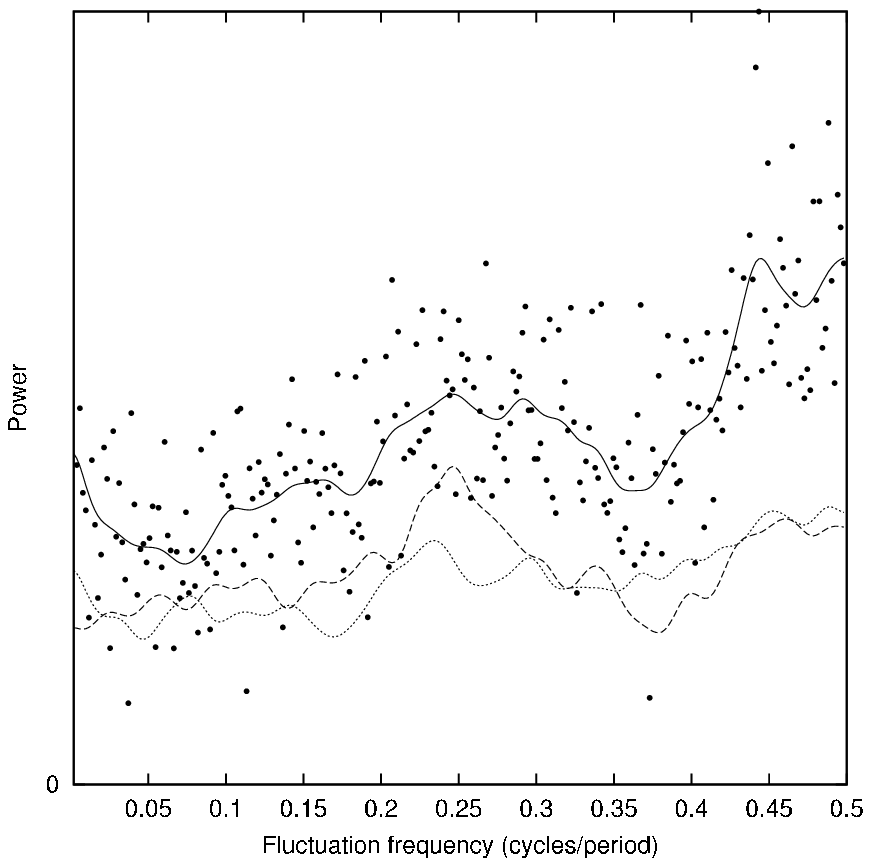}}
\caption{Fluctuation spectrum from the peak longitude bin (solid line)
and the bins immediately before (dashed) and after (dotted) it, for
PSR J2145$-$0750, at 1380 MHz. See caption for
Fig. \ref{fig:1022lrfs}.}
\label{fig:2145lrfs}
\end{figure}

In observations at 860 and 1380 MHz, this pulsar showed strong
intensity fluctuations, at some pulse longitudes exceeding $100\%$
(Fig. \ref{fig:2145modind}). Two 1380-MHz observations provided
sufficient sensitivity to detect weak quasi-periodicities around $\sim
0.22$ and $0.45$ cpp, we plot the result from the better observation
in Fig. \ref{fig:2145lrfs}. To check for stationary or drifting
modulation in the extended bridge and trailing component, we computed
the the 2DFS of this longitude interval. No evidence for
quasi-periodicity was found, but very broad features such as those
seen in the peak bins cannot be ruled out. In the 1380 MHz observation
we detected 490 individual pulses, the strongest of which had an
energy 3.7 times the mean. The profile formed by adding these pulses
was dominated by a narrow component around pulse longitude $215\degr$,
indicating that intensity enhancements in this component are not
accompanied by enhancements at other longitudes. The less sensitive
840 MHz observation yielded similar results for the 11 detected
pulses.

\section{Discussion}
\label{sec:discussion}
The results of this study  show that the emission from recycled
pulsars strongly varies on timescales comparable to the rotation
period.  Of the seven pulsars for which we were sensitive to
(aperiodic) modulation with $m > 0.5$, we detected such modulation in
all but one (PSR B1937+21). Quasi-periodicities also seem to be a
common feature of the fluctuation spectra, however features are
typically broad and often do not comprise the dominant contribution to
the observed modulation index. Such a broad, weak feature has also
been observed in PSR J0437$-$4715 \citep{vam98}. 

For PSR J1012+5307 and PSR J1518+4904 the presence of offset
components in the two-dimensional fluctuation spectra indicates that
the phase of the quasi-periodic modulations advances from the leading
edge to the trailing edge of the main profile component.  That is, the
components of the emission that are periodically modulated appear at
successively later longitudes. In addition to the results presented
here, we checked other observations and found similar behaviour.
These are the first ever detections of drifting subpulses in recycled
pulsars.  For the other pulsars the sensitivity to modulation at
longitudes away from the peak is probably too poor to have allowed
detection of similar phenomena, if present. For PSR J0437$-$4715 the
sensitivity achieved by \citet{vam98} appears more than sufficient to
have ascertained whether the modulation drifts, but unfortunately the
two-dimensional autocorrelation function was computed without prior
subtraction of the average profile, causing it to be dominated by this
component and potentially making any weak drift pattern invisible in
the plot. {\changed Determining whether or not the subpulse modulation drifts
would provide a good test of the model of \citet{gk97}, which predicts
stationary modulation.}

Despite the very different physical conditions in the polar cap
regions and magnetospheres of recycled pulsars versus ordinary
pulsars, the observed single-pulse phenomena are broadly consistent.
Specificially, like ordinary pulsars they tend to have modulation
indices of 0.5--1.5 (cf.\ \citealt{th71,bsw80,wab+86}) and non-white
fluctuation spectra (cf.\ \citealt{th71,brc75}). \citet{ran86}
distinguishes between modulation in so-called conal and core
components of the emission, noting that conal components frequently
show quasi-periodic modulation with periods of 2--10 pulses, which can
sometimes be highly coherent, while core components tend to possess
weak, broad low frequency excesses or peaks in their fluctuation
spectra\footnote{\citet{ran86} draws attention to a number of reports
of low-frequency quasi-periodicities in core-associated emission ($P_3
\sim 20$ pulses), however examination of the source literature (or at
least, those that published their spectra) indicates that with some
exceptions, most are just as consistent with a low frequency excess as
with quasi-periodicity.}.  We clearly do not detect any features of
high coherence in the recycled pulsars studied here, however it is
difficult to say what proportion of normal pulsars show them, for
studies have tended to favour pulsars known to possess them over those
with broad components that are more difficult to characterize. None of
the pulsars studied here appears to show a low-frequency excess
(except for the very low frequency scintillation components), perhaps
indicating that recycled pulsar emission is not related to core
emission in slow pulsars.

An apparent exception to the above generalizations is PSR B1937+21.
Our non-detection ($m < 0.26$) is consistent with the previously
measured modulation index of 0.18 \citep{jap01}\footnote{Since
\citet{jap01} dealt with baseband samples, the observed variance
included the radiometer noise of the pulsar signal itself. The
observed modulation index (for which they reported $1.032\pm 0.001$
and $1.034\pm 0.003$ in the main pulse and interpulse respectively) is
then given by $(1+2m^2)^{1/2}$.  In the use of integrated samples this
additional variance is reduced by the reciprocal of the time-bandwidth
product and is typically neglected.}.  The reported constancy of the
modulation index with longitude (except for the trailing edge, where
giant pulses occur) suggests scintillation as a possible origin for
the modulation, and \citet{jap01} show that the level of modulation is
consistent with the known properties of diffractive scintillation in
this source. \citet{jap01} use this result and the similarity of the
single-pulse ACF to the ACF of the average profile to argue that the
intrinsic emission of PSR B1937+21 is very stable, in contrast to all
previously studied pulsars.  However, \citet{wab+86} find that pulsars
they classified as ``core'' had low modulation indices, and indeed
their result for PSR B2053+36 resembles that of PSR B1937+21 not only
in the low value of modulation index but also in its constancy across
most of the profile and in the shape of the profile itself. The issue
could be resolved by examining the frequency structure of the
modulation seen in these pulsars.

In ordinary pulsars, the presence of drifting subpulses is usually
associated with a viewing geometry that traces the emission along a
quasi-transverse path near the edge of the beam
(e.g. \citealt{ran86}). This is consistent with the picture put forth by
\citet{rs75} that the pulsar beam consists of a circulating ring of
subbeams corresponding to such a pattern of spark activity on the
polar cap. The observation that drifting subpulse patterns are
coherent between different pulse longitudes and between observations
made simultaneously at different radio frequencies (i.e. that the
temporal period $P_3$ is independent of these; see
e.g. \citealt{tmh75}) lends strong support to this claim, as does the
recent apparent detection of modulation at the circulation period in
PSR B0943+10 \citep{dr01}. The presence of fluctuation components at
different temporal frequencies ought, in such a model, to obey a
mirror symmetry between pulse longitude pairs that sample the same
part of the rotating subbeam system. A clear example of this is PSR
B1237+25, where the relative importance of the (at least) three
modulation features shares the same approximate symmetry in longitude
as the average profile \citep{bac73}.

How PSR J1012+5307 fits into this picture, if at all, is unclear.
Taken at face value, the variation of peak fluctuation frequency with
pulse longitude implies that components Ia, Ib and II all correspond
to different (nested) rotating subbeam cones. The much higher
modulation index of component II versus component I would support this
assertion.  In this case one might expect that the emission from 290
to 340\degr\  corresponds to the complementary components
required to preserve symmetry mentioned above and ought to show a
progression from low to high fluctuation frequencies with longitude
(i.e. in reverse to that seen from Ia to Ib to II).

The detections of individual bright pulses made here are also broadly
consistent with the behaviour of ordinary pulsars, the pulse energy
distributions of which typically extend to a factor of a few higher
than the mean (e.g. \citealt{rit76}). A similar result for PSR
J0437$-$4715 was presented by \cite{jak+98}.  With the exception of
PSR B1937+21, for which giant pulses are already known to occur, and
PSR J0613$-$0200, for which the result of a single detection needs
further investigation, we did not detect any giant pulses. If any of
the pulsars observed had produced unresolved giant pulses with the
same energy probability distribution as that reported for PSR B1937+21
at 430 MHz \citep{cstt96}, we should have detected in the worst case
(PSR J0218+4232) $\sim 90$ of them. The results of \citet{kt00} appear
to indicate that for PSR B1937+21, giant pulses are less frequently
observed at higher frequencies, although they do not provide a fit to
the probability distribution. Our rough estimate based on Fig.~8 of
\citet{kt00}, given a mean flux density of 10.4 mJy (from the spectral
fit of \citealt{ffb91}) and a ratio of 0.68:0.32 in energy between the
main pulse and interpulse (measured from our observations) is that
giant pulses of energies exceeding a given multiple of the mean energy
occur of the order of 30 times less often at 1420~MHz than at 430~MHz.
Giant pulses with a similar probability distribution would have been
detected if present in any of the other pulsars observed.  This result
is particularly interesting for the first three pulsars listed in
Table \ref{tab:obs}, since among field recycled pulsars the inferred
magnetic field strengths at the light cylinder for these pulsars are
exceeded only by those of PSRs B1937+21 and B1957+20, making them
prime candidates for giant pulses according to \citet{cstt96}.

Other than the detections of giant pulses in PSR B1937+21 and PSR
B1812$-$24, we note that there exist in the literature largely
unexamined reports of strong pulses from PSR B1534+12
\citep{sb95,sal98}.  The pulse energy distribution clearly extends to
much higher energies than observed here for other recycled pulsars
(except PSR B1937+21), with the version given by \citet{sal98}
depicting the detection of pulses up to 17 times the mean energy, from
a sample of just 12000 pulses.  One wonders whether, were a data set
of a number of pulses comparable to those used in studies of PSR
B1937+21, PSR B1821$-$24 and the Crab pulsar analyzed ($\sim 10^7$),
sufficiently bright pulses would be received for them to be considered
``giant''. Such a result would be very interesting, since the inferred
magnetic field strength at the light cylinder for PSR B1534+12 is
low compared to most recycled and young pulsars ($\sim 1.6\times
10^3$~G, versus $\sim 10^6$~G for PSR B1937+21 and $\sim 9 \times
10^5$~G for the Crab pulsar). The suggestion of \citet{cstt96} that
this parameter is an indicator of giant pulse activity could be
strongly tested by studying strong pulses from PSR B1534+12.

Given the lack of a complete theory of pulsar emission, there are few
predictions that can be tested by the new observations presented here.
A recent attempt at a systematic prescription for the properties of
emission as a function of pulsar parameters for both normal and
millisecond pulsars is that of \citet{gs00}. They postulate that the
polar gap region is populated by a system of sparks that drift about
the center as a set of concentric ``rings''. Assuming that the spark
width and separation are both equal to the gap height, they define a
``complexity parameter'' ($a$), equal to the ratio of spark width to
the width of the polar cap, which they calculate as
\begin{equation}
a = 5 \left(\frac{\dot{P}}{10^{-15}}\right)^{2/7} 
\left(\frac{P}{1\;{\rm s}}\right)^{-9/14},
\end{equation}
\noindent where $P$ is the pulsar spin period. Since the space between
sparks is equal to their width, the total number of sparks across the
cap is $\sim a$, and the number of nested cones of emission (excluding
the centre spark) is $n \simeq (a-1)/2$. They show that the complexity
parameter is correlated with profile morphology, with core single
profiles having the highest complexity parameters, followed by triple
profiles and then multiple and conal single profiles.  This leads
them to suggest that core single profiles, and the central
components of triple profiles, are the result of emission from a large
number of narrow cones, broadened by the finite beam-width of the
elementary emission {(\changed$2/\gamma$, where $\gamma$ is the Lorentz factor
of the emitting particles). This explains why their profile morphology
is simple, and makes the prediction that the modulation index of such
pulsars should be small. }

\begin{figure}
\resizebox{\hsize}{!}{\includegraphics{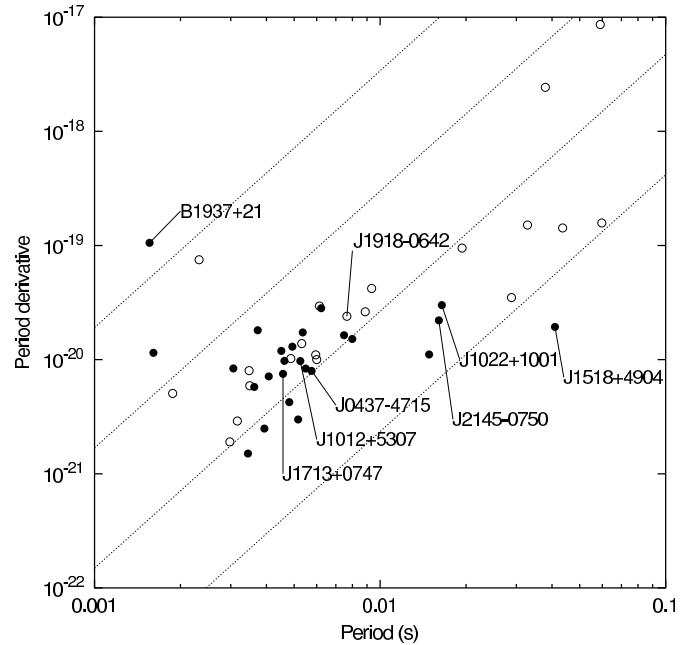}}
\caption{Period vs Period derivative for 46 recycled pulsars in the
Galactic disk. The solid points have been corrected for acceleration
effects, the unfilled points have not. The dotted lines correspond to
lines of equal value in the complexity parameter of \citet{gs00},
$a=20,10,5,2.5$ (top to bottom).}
\label{fig:ppdot}
\end{figure}

In Figure \ref{fig:ppdot} we plot lines of constant complexity
parameter along with the period and intrinsic period derivative for 46
field pulsars presumed to be recycled ($P < 0.1$ s, $\dot{P} <
10^{-15}$). We used the published proper motions of \citet{tsb+99} and
\citet{lcw+01}, plus a new measurement for PSR J1518$+$4904
($\mu=7.5\pm 0.7$ mas yr$^{-1}$; Stappers unpublished), in combination
with the Galactic potential model of \citet{kg89} to correct the
observed period derivatives for the effects of acceleration with
respect to the line of sight \citep{dt91}. These points are shown as
solid circles, while uncorrected points are not filled. The sources
for which the modulation properties are now known are labelled.
{\changed For most sources the model predicts 1--3 cones of emission,
implying complex profiles with significant modulation, as
observed. The exception to this statement is PSR B1937+21, with a
large complexity parameter of $a\simeq 23$.  This pulsar is also
exceptional for its low modulation index and simple profile
morphology, just as expected for a large number of unresolved
cones. We have already drawn an analogy between PSR B1937+21 and PSR
B2053+36, and it is interesting to note that the complexity parameter
of the latter is also quite high: $a\simeq 10$.  Indeed,
\citet{wab+86} found that not only PSR B2053+36, but all of the
core-dominated pulsars in their sample have low modulation indices,
in excellent agreement with the assertion, made by \citet{gs00} on the
basis of the tendency for core pulsars to have large complexity
parameters, that core components are the sum of numerous
$1/\gamma$-smeared components.}

We have shown that the detection and characterization of
pulse-to-pulse intensity fluctuations in millisecond pulsars is
possible using existing instruments. The acquistion of similar
information about southern recycled pulsars and those accessible from
Arecibo should expand the available sample and make stronger
generalizations about the population possible. With the possible
future availability of very large radio telescopes such as LOFAR and
SKA both the number of recycled pulsars within the reach of study, and
the depth of detail to which they could be studied will be greatly
enhanced and hopefully provide considerable constraining information
on recycled pulsar polar caps and magnetospheres.

\acknowledgements
We thank R.~Strom, A.~Lommen, R.~Ramachandran and M.~Kramer for allowing
us to use data originally recorded for other projects. RTE is supported
by a NOVA fellowship. The WSRT is operated by ASTRON with financial
support from the Netherlands Organisation for Scientific Research
(NWO).


\end{document}